# Simulation and experiments of Stacks of High Temperature Superconducting Coated Conductors Magnetized by Pulsed Field Magnetization with Multi-Pulse Technique

Shengnan Zou[1], Víctor M. R. Zermeño[1], A. Baskys[2], A. Patel[2], Francesco Grilli[1], B. A. Glowacki[2]

[1] *Institute for Technical Physics, Karlsruhe Institute of Technology, 76344 Eggenstein-Leopoldshafen, Germany*
E-mail: shengnan.zou@kit.edu

2 *Department of Materials Science and Metallurgy, University of Cambridge, Cambridge CB3 0FS, U.K.*

## Abstract

High temperature superconducting (HTS) bulks or stacks of coated conductors (CCs) can be magnetized to become trapped field magnets (TFMs). The magnetic fields of such TFMs can break the limitation of conventional magnets (<2 T), so they show potential for improving the performance of many electrical applications that use permanent magnets like rotating machines. Towards practical or commercial use of TFMs, effective *in situ* magnetization is one of the key issues. The pulsed field magnetization (PFM) is among the most promising magnetization methods in virtue of its compactness, mobility and low cost. However, due to the heat generation during the magnetization, the trapped field and flux acquired by PFM usually cannot achieve the full potential of a sample (acquired by the field cooling or zero field cooling method). The multi-pulse technique was found to effectively improve the trapped field by PFM in practice. In this work, a systematic study on the PFM with successive pulses is presented. A 2D electromagnetic-thermal coupled model with comprehensive temperature dependent parameters is used to simulate a stack of CCs magnetized by successive magnetic pulses. An overall picture is built to show how the trapped field and flux evolve with different pulse sequences and the evolution patterns are analyzed. Based on the discussion, an operable magnetization strategy of PFM with successive pulses is suggested to provide more trapped field and flux. Finally, experimental results of a stack of CCs magnetized by typical pulse sequences are presented for demonstration.

## 1. Introduction

High temperature superconducting (HTS) bulks (REBaCuO, $MgB_2$ or iron pnictides) or composite bulks (stacks of REBaCuO coated conductors (CCs)) can be magnetized to become trapped field magnets (TFMs) by trapping persistent currents. Such TFMs can generate a magnetic field which breaks the limitation of conventional permanent magnets (<2 T). These impressive world-record trapped fields are 17.6 T at 26 K for bulks [1] and 7.92 T at 4.2 K for stacks of CCs [2][3]. The remarkable high field of TFMs is promising to increase the power/torque density of rotating machines [4] and improve the performance of other applications, such as magnetic flywheels [5], magnetic separator [6] and nuclear magnetic resonance magnets [7].

Towards practical use or commercialization of TFMs in electrical applications, cheap and compact magnetization of TFMs is one of the key issues. The remarkable world-record trapped fields were achieved by the field cooling (FC) or zero field cooling (ZFC) method, where a quasi-static background field of the magnitude one (FC) or two (ZFC) times of the final trapped fields [8] needs to be provided. Superconducting magnets with huge power sources are needed for such processes, which is unacceptable for practical applications. The so-called pulsed field magnetization (PFM) is a promising substitute to achieve *in situ* magnetization. Instead of providing a quasi-static high field, a pulsed magnetic field [9] of milliseconds is used to realize compact, movable and cost-effective magnetization. However, the trapped field obtained by PFM is generally lower than that by FC or ZFC, especially at lower temperatures [10]. This is due to the fast flux motions during PFM which generate substantial heat and increase the temperature [11][12]. Effort has been devoted to optimize the PFM process to improve the trapped field, which can be found in the recent review [10].

It has been experimentally proved that the multi-pulse technique can improve the trapped field of HTS bulks by PFM effectively. The multi-pulse techniques include successive pulsed-field application (SPA), iteratively magnetizing pulsed field method with reducing amplitude (IMRA), and multi-pulse technique with step-wise cooling (MPSC) [13],[14],[15],[16]. The increase of the trapped field by successive pulses is understood as a result of reducing temperature rise due to the existence of currents produced by previous pulses. However, little numerical simulation of multi-pulse technique has been reported and they were carried out only for bulks [17],[18],[19],[20],[21]. The mechanism of the improvement in the trapped field and flux by multi-pulse technique is not clear yet. The applied pulse sequence used in experiments is usually determined empirically.

In this work, a systematic study on the PFM with SPA and IMRA is presented. Stacks of HTS CCs magnetized by multi-pulse technique at a fixed temperature are simulated. The stack is chosen for simulation for the increasing interest in it [22],[23],[24],[25]. Compared to pure HTS bulks, stacks of HTS CCs have several advantages: the HTS tapes are more commercialized and the stacks are relatively cheaper; the mechanical strength of stacks is better thanks to the metal substrates; and the shapes of stacks are more flexible to adapt to different configurations of applications [25]. Furthermore, HTS tapes are easier to characterize reliably to allow trustable modelling [21]; and they are more uniform without obvious grain boundaries. Considering that the non-linear *E-J* properties of bulks and tapes are generally similar, the derived patterns and conclusions may apply to HTS bulks as well.

The aim of this work is to understand the mechanism of the multi-pulse technique and suggest an optimal applied pulse sequence at a fixed temperature (corresponding to SPA and IMRA technique) by numerical simulation. The simulation provides a theoretical direction for practice while saving the considerable time or costs required by experiments to try out numerous pulse sequences. The model is a 2D electromagnetic-thermal coupled model using finite element method (FEM) describing an infinite long stack. It considers realistic structures of HTS CCs with all composite materials. The anisotropic magnetic field and temperature dependent critical current density of CCs and other temperature dependent parameters of different materials are considered comprehensively. The stack magnetized by three successive pulses of different combinations of pulse magnitudes is simulated to present a complete evolution map of the trapped field and flux by different magnetization sequences. Based on the derived patterns, operable optimal applied pulse sequences are suggested to provide more trapped field and flux. Finally, experiments are carried out on a stack of CCs magnetized by different pulse sequences for verification.

2. **Model description**

The model is a 2D electromagnetic-thermal coupled model using the finite element method (FEM) implemented in Comsol 5.2. The details of the model can be found in [26]. The electromagnetic part of the model is described by the so-called ***H***-formulation, which is a well demonstrated approach for evaluating the AC loss of HTS [27],[28],[29],[30]. The Maxwell's equations are solved for variables $H_x$ and $H_y$ in the 2D planar model with the ***E-J*** power law [31],

$$\mu \cdot \frac{\partial H_x}{\partial t} + \frac{\partial E_z}{\partial y} = 0,$$

(1a)

$$\mu \cdot \frac{\partial H_y}{\partial t} - \frac{\partial E_z}{\partial x} = 0,$$

(1b)

$$E_z = E_c \left( \left| \frac{J_z}{J_c(\boldsymbol{B}, T)} \right| \right)^n,$$

(1c)

where $E_c$ equals $10^{-4}$ V m$^{-1}$.

In this model, anisotropic magnetic field and temperature dependent $J_c(\boldsymbol{B},T)$ is considered using the elliptical equation [32],

$$J_c(B,T) = J_{c0}(T)\left(1 + \frac{\sqrt{(k(T)B_x)^2 + B_y^2}}{B_0(T)}\right)^{-b(T)},$$

(2)

which contains four temperature dependent parameters: $J_{c0}(T)$, $k(T)$, $B_0(T)$ and $b(T)$. The equation is generally valid to describe the main features of the anisotropic critical current density of commercial 2G HTS tapes in the absence of artificial pinning. The parameter $k(T)$ indicates the anisotropy of magnetic field dependence of the critical current density. $B_0(T)$ describes the decreasing speed of the critical current density with the magnetic field. These parameters are temperature dependent. In this work, the values are derived from the lift factor measurement data of YBCO tapes (SCS4050) by Superpower Inc.[33]. The parameters are provided in [26].

The electromagnetic model is coupled with the heat transfer equation in solids,

$$\rho C_p \frac{\partial T}{\partial t} = \nabla \cdot (\lambda \nabla T) + Q,$$

(3a)

$$Q = E_z J_z,$$

(3b)

where $\rho$, $C_p$ and $\lambda$ are the mass density, heat capacity and thermal conductivity, respectively. In this work, the temperature dependence of $C_p$ and $\lambda$ are considered. Taking into account of such temperature dependence is necessary, because during the PFM process the temperature increases by tens of kelvins and the values of thermal parameters change substantially.

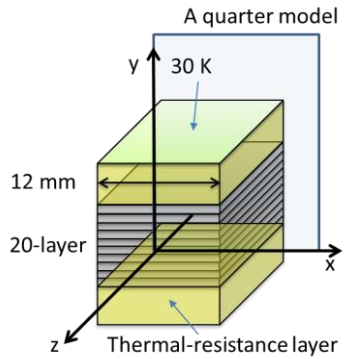

Fig. 1 The geometry of the model. A quarter of the cross-section of the stack is simulated. The plot is not to scale.

The geometry of the model is shown in Fig. 1. The realistic structure of laminated YBCO tapes (Superpower Inc.[34]) is considered, including the Hastelloy substrates, silver overlayers, copper stabilizers and YBCO layers. Only the 0.2 μm thick buffer layers are neglected: due to their extreme thinness, they do not modify significantly the thermal exchange between the other layers. All the layers, including the superconducting one, are simulated with their real dimensions. Comprehensive temperature dependent thermal properties of different composite materials are considered. The thermal boundary is set by putting a thermal-resistance layer between the tape surface and a 30 K constant temperature boundary. In the present model, the layer is 1 mm thick with $\lambda = 0.1$ W/(m·K). The layer's thickness and thermal conductivity can be adjusted to represent different the cooling powers. Full sets of parameters can be found in [26]. Considering symmetry, only a quarter of the cross-section of the stack is simulated to save computing time.

The stack is magnetized by three successive pulses as shown in Fig. 2. The amplitudes of the three pulses can be different. The shape of each pulse follows the equations (Pulse 1):

$$B_{app} = B_{app1} \sin^2\left(\frac{\pi t}{2\tau}\right), 0 \leq t < \tau$$

(4a)

$$B_{app} = B_{app1} \cos^2\left(\frac{\pi(t-\tau)}{10\tau}\right), \tau \leq t \leq 6\tau$$

(4b)

where $\tau = 10$ ms. These equations have smooth transitions at the beginnings and peaks, which lead to better convergence. The pulse takes 10 ms to ramp to the peak amplitude and 50 ms to damp to zero. After each pulse, the stack relaxes for 20 s and the overall temperature decreases to 30 K due to the cooling. Then, the next pulse is applied. The trapped field or flux after each pulse is measured after relaxing (at Rm1, 2 and 3).

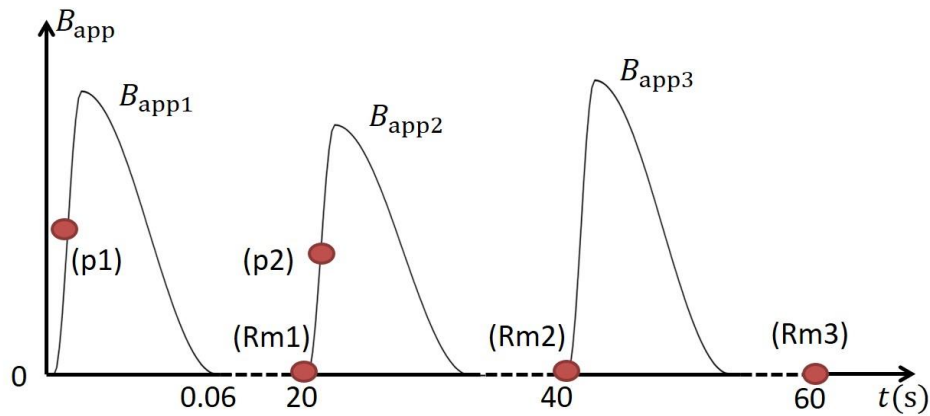

**Fig. 2** Temporal evolution of the three applied pulses. The marked time points will be used for later discussion.

## 3. Results and discussion

In this work, the trapped fields ($B_y$) are measured 0.8 mm above surface center of the stack. The trapped fluxes are defined as the integration of the trapped field 0.8 mm above the stack surface from x=0 mm to x=5 mm, which includes mostly regions of positive remnant fields for the desired cone-shaped field distributions [10]. The exact values of trapped flux depend on the chosen region for integration and the distance between the measurement positions and the stack surface considering specific applications. The definition of trapped flux influences the exact values, but it doesn't change the tendency and analysis in this work. The trapped field and flux obtained by FC or ZFC in the stack are 1.377 T and 3.67 mWb/m, respectively. They are calculated with static models by assuming that the current density everywhere inside the stack equals the critical current density. Detailed descriptions of the models can be found in [35].

The stack is magnetized by three successive pulses as shown in Fig. 2. There are numerous combinations of the amplitudes of the three pulses. The discussion is arranged to clarify the patterns of how the trapped field and flux evolve after each pulse to find out the optimal applied pulse sequences.

### 3.1 Pulse 1

The trapped field and flux with the amplitudes of Pulse 1 are shown in Fig. 3 and Fig. 4, respectively. The trapped field and flux are measured 20 s after the pulse (Rm1 in Fig. 2) after relaxation, when the temperature has returned to 30 K. Only representative data points are shown to avoid confusion by massive data, especially after they evolve with the number of pulses in later sections. For the trapped field (Fig. 3), results are given for $B_{app1}$ = 2.2 T, 2.6 T and 3 T ($B_{app1}$ = 1.8 T produces only 0.312 T trapped field, so it is excluded from the figure to avoid unnecessary plotting); for the trapped flux (Fig. 4), results are given for $B_{app1}$ = 1.8 T, 2.2 T, 2.6 T and 3 T.

As shown in Fig. 3, there is an optimal applied field (2.6 T in Fig. 3), which gives the maximum trapped field after Pulse 1. If the applied pulse is too small, the sample is not fully magnetized (namely under-magnetized); however, if the applied pulse is too large, excessive heat is generated and the critical current density will be over reduced (namely over-magnetized). This can be understood better by looking at the normalized current density distribution ($J/J_c$) in Fig. 5. When under-magnetized, the current density is higher, but there are still large regions with negative currents; when over-magnetized, the whole stack is filled with positive currents as desired, but the current density is lower due to excessive heat generation. The optimal magnetized situation is a compromise when there are still negative currents but the overall current density is relatively higher. This result is consistent with previous literature, for both numerical and experimental results [10],[24].

The trapped flux shows a similar pattern with an optimal applied field giving the maximum trapped flux. However, the optimal applied field (2.2 T) giving the maximum trapped flux is smaller than that (2.6 T) giving the maximum trapped field. The trapped field is the single field value in the center, while the trapped flux is an integration of the fields along the interested region. For the former, more penetration (larger applied field) of the positive currents towards the center is more beneficial, because more positive currents are closer to the interested point and contribute more fields according to the Biot-Savart law. For the latter, larger positive current density on the periphery region contributes more to the total flux, so less penetration (smaller applied field) is preferred. The following discussion will focus on looking for the optimal applied pulse sequence that can produce both the maximum trapped field and flux.

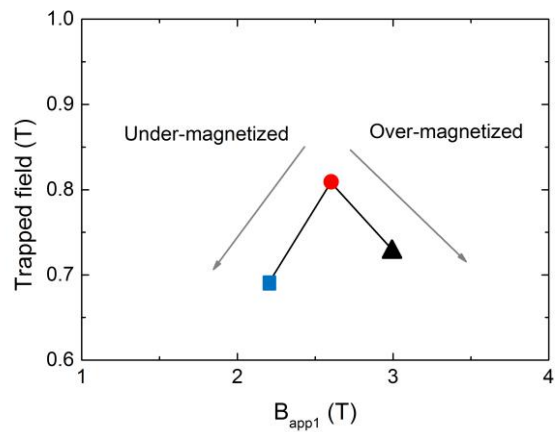

Fig. 3 The trapped field with the amplitude of Pulse 1 (at time Rm1 in Fig. 2).

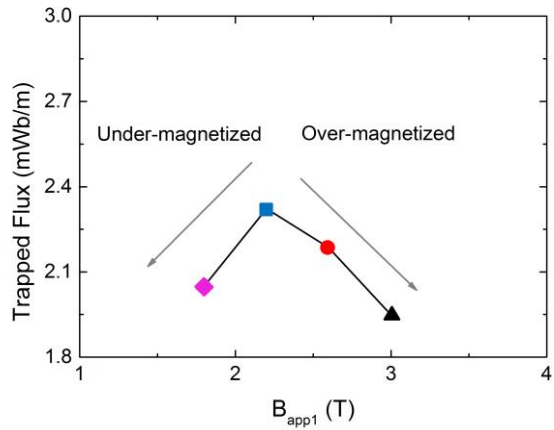

Fig. 4 The trapped flux with the amplitude of Pulse 1 (at time Rm1 in Fig. 2).

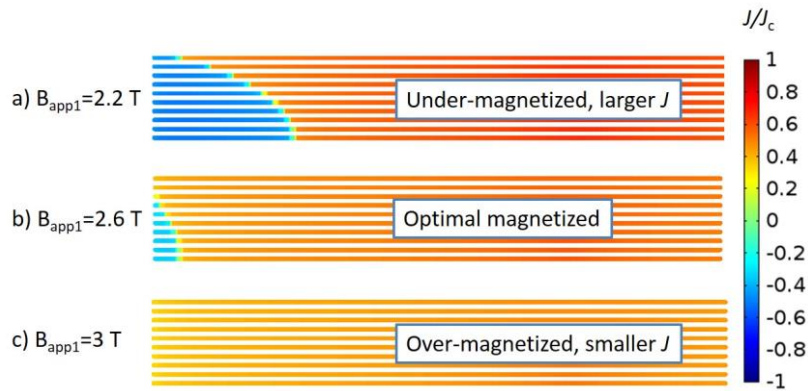

Fig. 5 The normalized current density distribution after Pulse 1 (at time Rm1 in Fig. 2). Only a quarter of the cross-section of the stack is shown, as marked in Fig. 1. The thickness of the HTS layer is exaggerated for visualization.

### 3.2 Pulse 2

After Pulse 2, the trapped field and flux evolve from the values after Pulse 1 into new values. The trapped field and flux with the amplitude of Pulse 2 (at Rm2 in Fig. 2) are plotted in Fig. 6 and Fig. 7, respectively. Figuratively speaking, each point in Fig. 3 and Fig. 4 grows into a line in Fig. 6 and Fig. 7, respectively.

As shown in Fig. 6, there is an optimal applied field for Pulse 2 which gives the maximum trapped field. Otherwise the stack is either under-magnetized or over-magnetized. The optimal applied fields (2.6 T) are the same for the three lines and are also the same with the previous optimal applied field (2.6 T) of Pulse 1. For the trapped flux in Fig. 7, there is also an optimal applied field which gives the maximum trapped flux for each line. The optimal applied fields are different for the four lines (1.8 T or 2.2 T). So far, the maximum trapped field is reached by the pulse sequence $B_{app1} = 2.6$ T, $B_{app2} = 2.6$ T; the maximum trapped flux is reached with $B_{app1} = 2.6$ T, $B_{app2} = 2.2$ T.

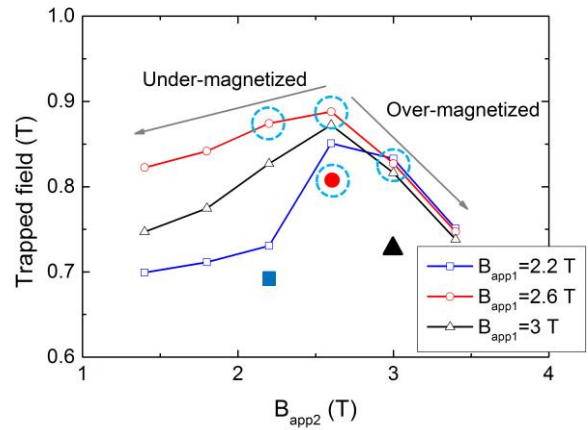

Fig. 6 The trapped field with the amplitude of Pulse 2 (at time Rm2 in Fig. 2). The three full symbols are the trapped fields after Pulse 1 (Fig. 3), which grow into the three lines. The marked points will be used in Fig. 8.

The normalized current density distributions after Pulse 2 (Rm2 in Fig. 2) for typical cases (marked points in Fig. 6 and Fig. 7) are shown in Fig. 8. After the second pulse, the current distributions resulted from Pulse 1 evolve into the three situations:

a. Smaller second pulse ($B_{app2} = 2.2$ T $< B_{app1} = 2.6$ T). The penetration is less. Only the currents on the periphery of the stack are influenced. The current density on the periphery is enhanced, because the applied pulse is smaller and the temperature is lower compared to that during Pulse 1. The trapped field is always improved. Yet if Pulse 2 is too small, the improvement will be less.
b. Larger second pulse ($B_{app2} = 3$ T $> B_{app1} = 2.6$ T). The previous penetration will be buried. The trapped field can be improved or reduced.
c. Equal second pulse ($B_{app2} = 2.6$ T $= B_{app1} = 2.6$ T). The penetration is less compared to the previous. The current density on the influenced part is improved. The magnetic field and flux line distribution at the same time (5 ms) of the ascending stage of the first and second pulse (time p1 and p2 in Fig. 2) are compared in Fig. 9. During the magnetization by Pulse 2, the existing currents (resulted from Pulse 1) in the stack expel the flux lines, so the penetration is less compared to Pulse 1, the magnetic flux density is generally lower and the temperature rise is lower. As a result, repeating the same pulse always improves the trapped field.

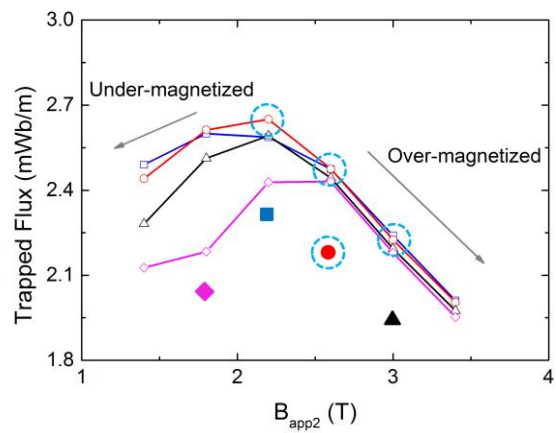

Fig. 7 The trapped flux with the amplitude of Pulse 2 (at time Rm2 in Fig. 2). The four full symbols are the values after Pulse 1 (Fig. 4), which grow into the four lines after Pulse 2. The marked points will be used in Fig. 8.

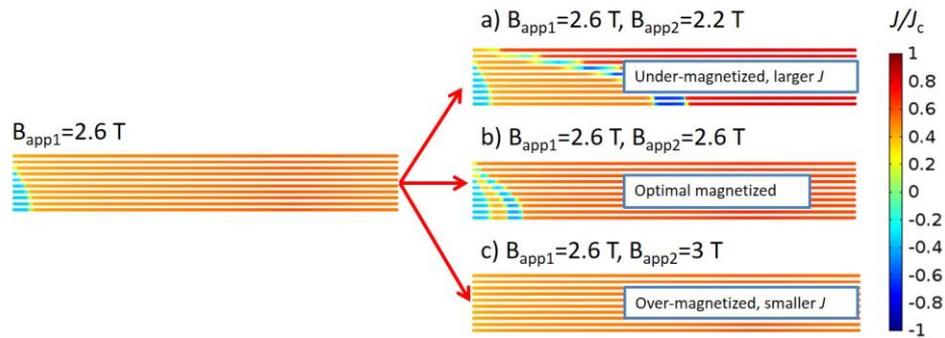

Fig. 8 The normalized current density distribution (marked points in Fig. 6 and Fig. 7) after Pulse 2 (at time Rm2 in Fig. 2). Only a quarter of the cross-section of the stack is shown, as marked in Fig. 1. The thickness of the HTS layer is exaggerated for visualization.

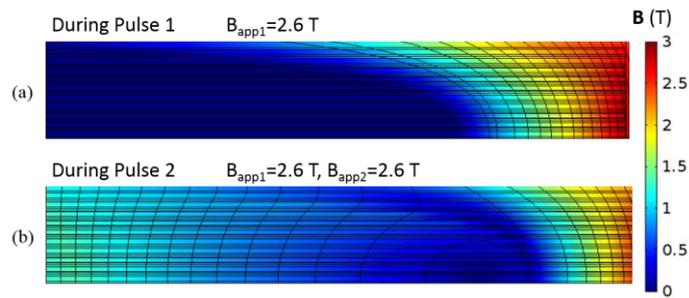

Fig. 9 The magnetic field and flux line distribution during the ascending stage of Pulse 1 and Pulse 2 (p1 and p2 in Fig. 2).

### 3.3 Pulse 3

Similar to the evolution from Pulse 1 to Pulse 2, after Pulse 3 the trapped field and flux evolve from the values after Pulse 2 into new values. To provide an easy way to follow the analysis, an evolution tree describing the trapped field with the pulse number is shown in Fig. 10. The trapped field evolves with the magnetization of successive pulses. The evolution follows different routes, which depends on the amplitude of each pulse.

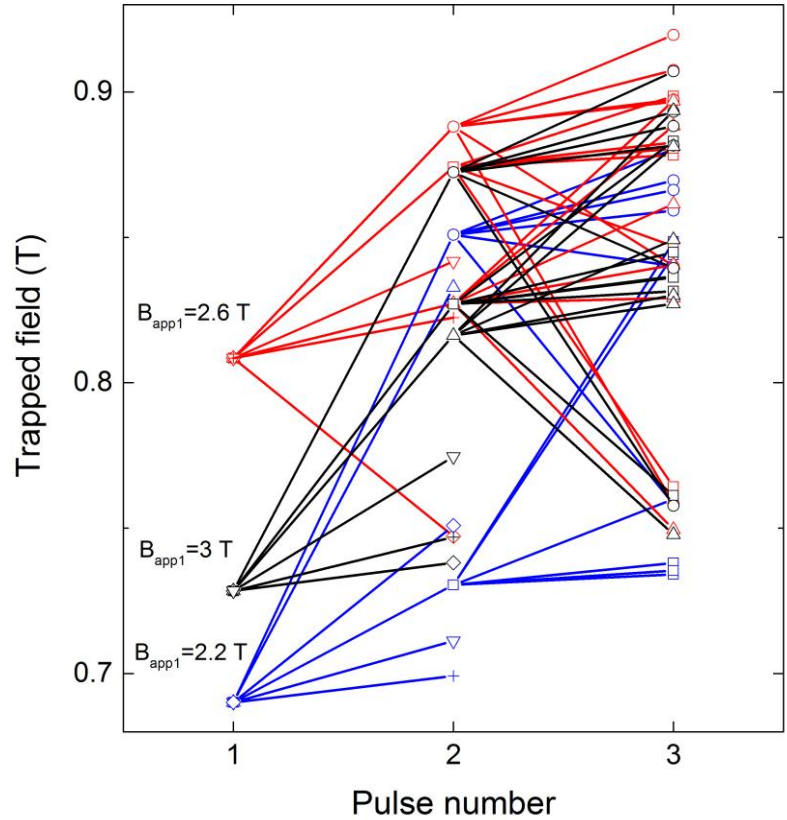

Fig. 10 The evolution tree: the trapped field with the pulse number. The trapped field evolves with pulse number following different routes, depending on the amplitude of each pulse.

Again, each point in Fig. 6 and Fig. 7 grows into a line in Fig. 11 and Fig. 12, which show the trapped field and flux with the amplitude of Pulse 3 (Rm3 in Fig. 2), respectively. To avoid the confusion caused by too many lines, only selected data are plotted. For the trapped field, nine points ($B_{app1}$ = 2.2 T, 2.6 T and 3 T; $B_{app2}$ = 2.2 T, 2.6 T and 3 T) in Fig. 6 are chosen to generate nines lines in Fig. 11. To provide better visualization, the nine lines are divided into three sub figures depending on $B_{app2}$. And different colors or symbols distinguish different $B_{app1}$. The trapped fields after previous pulses are kept for reference (full symbols represent Pulse 1, empty symbols with dash lines represent Pulse 2). Taking the black solid line in Fig. 11(a) as an example, the trapped field is the black triangular full symbol at Rm1 ($B_{app1}$ = 3 T), then it becomes the strengthened black triangular empty symbol at Rm2 ($B_{app2}$ = 2.2 T) and finally becomes the black solid line. Similarly, Fig. 12 shows the trapped flux evolved from 16 points ($B_{app1}$ = 1.8, 2.2 T, 2.6 T and 3 T; $B_{app2}$ = 1.8, 2.2 T, 2.6 T and 3 T) in Fig. 7 with the same visualization technique explained above.

As shown in Fig. 11, the trapped field first increases and then decreases with the amplitude of Pulse 3, similar to the situation for Pulse 1 and Pulse 2. The trapped fields are generally improved after the application of Pulse 3. Taking the marked points in Fig. 11 and Fig. 12 as an example, the normalized current density distributions are plotted in Fig. 13 for analyzing the penetration process. After Pulse 3, the current distribution resulted from previous pulses ($B_{app1}$ = 2.6 T; $B_{app2}$ = 2.6 T) evolves into three situations again (similar to the discussion in Section 3.2):

a. Smaller third pulse ($B_{app3}$ = 2.2 T < $B_{app2}$ = 2.6 T). The penetration is even less. Only the currents on the periphery of the stack are influenced. The current density on the periphery is enhanced, because the applied pulse is smaller and the temperature is lower. The trapped field is always improved. Yet if Pulse 3 is too small, the improvement will be less.

b. Larger third pulse ($B_{app3} = 3$ T $> B_{app2} = 2.6$ T). The previous penetration (by Pulse 1 and 2) will be buried. The trapped field can be improved or reduced.
c. Equal third pulse ($B_{app3} = 2.6$ T $= B_{app2} = 2.6$ T). By repeating the same amplitude for the third time, the penetration is saturated. The trapped field is not improved anymore (or insignificant improvement).

The maximum trapped field achieved after Pulse 3 is obtained by using the pulse sequence $B_{app1} = 2.6$ T, $B_{app2} = 2.6$ T and $B_{app3} = 2.2$ T. Combined with the analysis above, the trapped field is improved by magnetizing the stack optimally from the center to the periphery of the stack section by section. The center of the stack demands large pulse to be magnetized, but the large pulse will also increase the temperature and sacrifice the periphery part. After the magnetization of the central part, smaller pulses should be used to magnetize the outer section again without influencing the central part. And this can be done successively until all parts are optimally magnetized.

With the above discussion about penetration, the evolution patterns in Fig. 11 can be understood. Fig. 11(a) has the optimal applied fields at $B_{app3} = 2.6$ T and shows characteristic points at $B_{app3} = 2.2$ T. Below 2.2 T the trapped field is not changing much because $B_{app3}$ is too small which only influences the periphery. Above 2.2 T, the trapped field can be increased at 2.6 T by burying part of the old penetration ($B_{app2} = 2.2$ T) and make more use of the relatively central part. Even larger Pulse 3 (3 T) will bury the old penetration and may reduce the trapped field due to more heat generation. At larger Pulse 3 amplitudes, the three lines are overlapping, because all old penetrations by Pulse 1 and 2 are buried and erased. Fig. 11(b) has the optimal applied fields at $B_{app3} = 2.2$ T, because the inner part of the stack has already been optimally magnetized by Pulse 1 and 2. More trapped field can be acquired by making use of the relatively outer part by applying a smaller pulse. So the optimal applied field is shifted to smaller values. Fig. 11(c) has the optimal applied fields at $B_{app3} = 2.6$ T. Pulse 2 (3 T) was too large and it has erased Pulse 1, so Pulse 3 can improve the trapped field by magnetizing the central part of the stack optimally with a smaller pulse (2.6 T).

Generally speaking, there is an optimal applied field for each applied pulse which gives the maximum trapped field. The optimal applied field will shift to a smaller value after the saturation by the previous optimal applied field, which has optimally magnetized the relatively central part of the stack. The saturation is achieved by applying the previous optimal applied field at least once. So Fig. 11(a) and (c) still have the optimal applied field 2.6 T, which is the same as that of Pulse 1 and 2; while Fig. 11(b) was saturated by the previous optimal applied field (2.6 T) of Pulse 1 and 2 and has a smaller optimal applied field 2.2 T for Pulse 3.

For the trapped flux shown in Fig. 12, the patterns are similar with the trapped field, but the optimal applied fields are different. The reason is that the periphery part of the stack contributes more to the total flux compared to the central part. The optimal applied field is fast shifting to smaller values to make more use of the periphery parts of the stack. Fig. 12(a) is of the same pattern with Fig. 11(a); Fig. 12(b) with Fig. 11(b); Fig. 12(c) and (d) with Fig. 11(c). Similarly, to achieve more trapped flux, the stack should be magnetized section by section from the center to the periphery to make optimal use of each part of the stack.

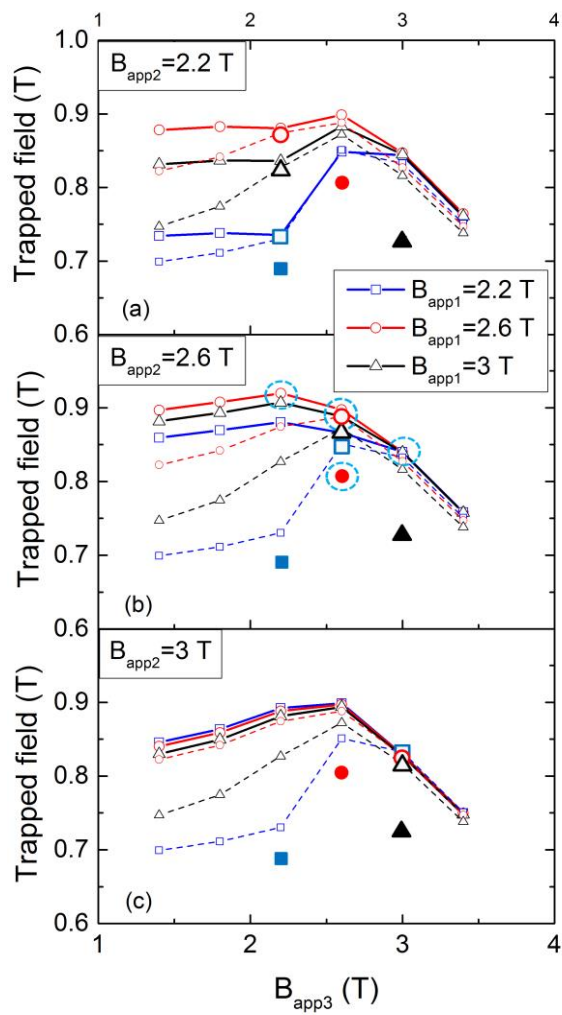

Fig. 11 The trapped field with the amplitude of Pulse 3 (at time Rm3 in Fig. 2). The three full symbols are the values after Pulse 1 (Fig. 3), and the dashed lines are the values after Pulse 2 (Fig. 6). See text for detailed explanations. The marked points will be used in Fig. 13.

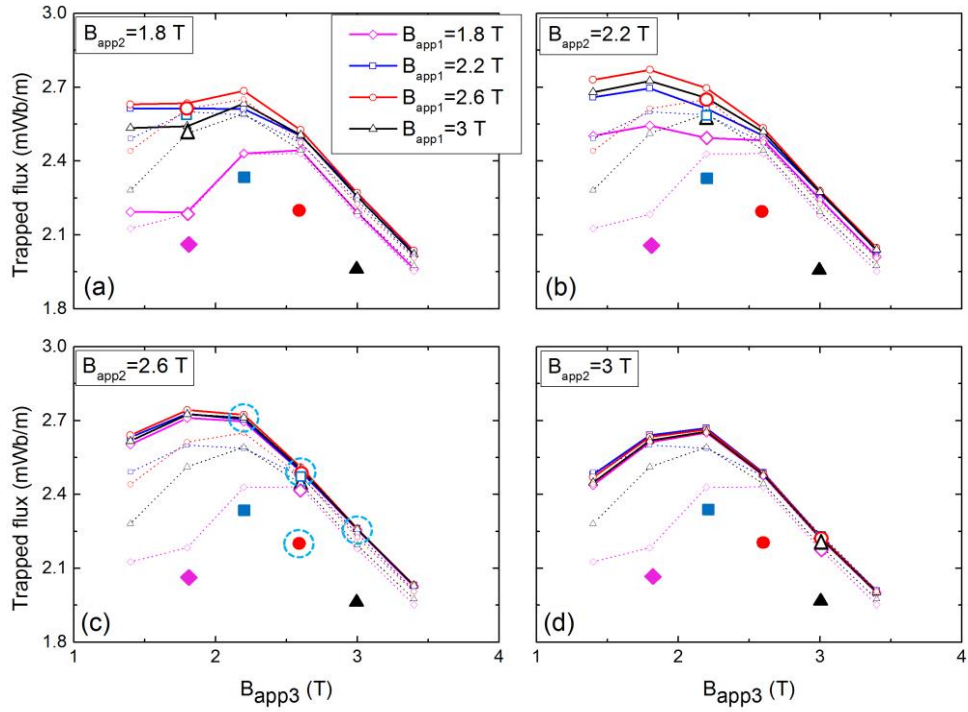

Fig. 12 The trapped flux with the amplitude of Pulse 3 (at time Rm3 in Fig. 2). The three full symbols are the values after Pulse 1 (Fig. 4), and the dashed lines are the values after Pulse 2 (Fig. 7). See text for detailed explanations. The marked points will be used in Fig. 13.

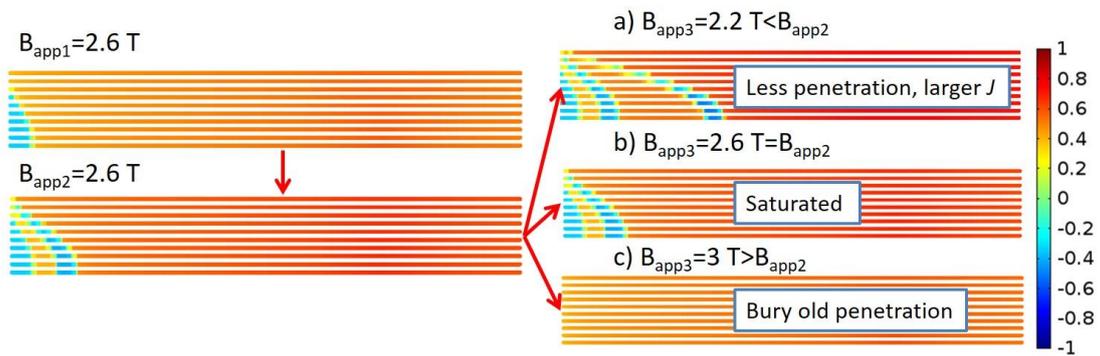

Fig. 13 The normalized current density distributions after Pulse 3 (at time Rm3 in Fig. 2). Only a quarter of the cross-section of the stack is shown, as marked in Fig. 1. The thickness of the HTS layer is exaggerated for visualization.

### 3.4 Summary of patterns and the optimal sequence

Summarizing the analysis of Section 3.1 to 3.3, the conclusions are:

a. For each applied pulse, there is an optimal applied field which gives the maximum trapped field and trapped flux. Otherwise the stack is either under-magnetized or over-magnetized. The optimal applied field for the trapped flux is usually equal or smaller than that for the trapped field.
b. Equal successive pulses can always improve the trapped field or flux, but the improvement will saturate soon.
c. Smaller successive pulses can always improve the trapped field or flux by making more use of the relatively outer part of the stack. If the successive pulse is too small, the improvement will be less.
d. Larger successive pulses will bury the old penetration. If it is too large, the trapped or flux will can be decreased due to excessive heat generation.
e. The optimal magnetization strategy is by optimally magnetizing the stack section by section from the center to the periphery of the stack with pulses of equal or descending amplitudes, so that each section of the stack can be used optimally.

According to these findings, an optimal applied pulse sequence is qualitatively plotted in Fig. 14. When a sample is given, the optimal amplitude of the starting pulse is probably unknown. The practical approach is to apply a relatively large starting pulse ($B_{m1}$) to over-magnetize the stack. Then repeat the same amplitude (Pulse 2, 3 in Fig. 14) until saturation is observed. Saturation means that the trapped field and flux are not increasing significantly any more. After the saturation, a smaller pulse (Pulse 4, $B_{m2}$) should be applied. Theoretically, the reduction in the amplitude of the pulse is infinitesimal. Then repeat the same amplitude again (Pulse 5, 6) until saturation is observed again. Then, Pulse 7 of an even smaller amplitude $B_{m3}$ is applied. And the same will be done on and on. In this way, the stack is over-magnetized at the beginning, until the optimal amplitude for the central part is applied. And then the applied amplitudes keep repeating or decreasing and the stack is optimally magnetized from the center to the periphery section by section.

In practice, it is favorable to magnetize a sample with less time or a smaller number of pulses. Considering that the repeated pulses barely improve the trapped field and flux any more after the second pulse, only the pulses with descending amplitudes are necessary, as shown in Fig. 15. This will be further justified in Section 3.5.

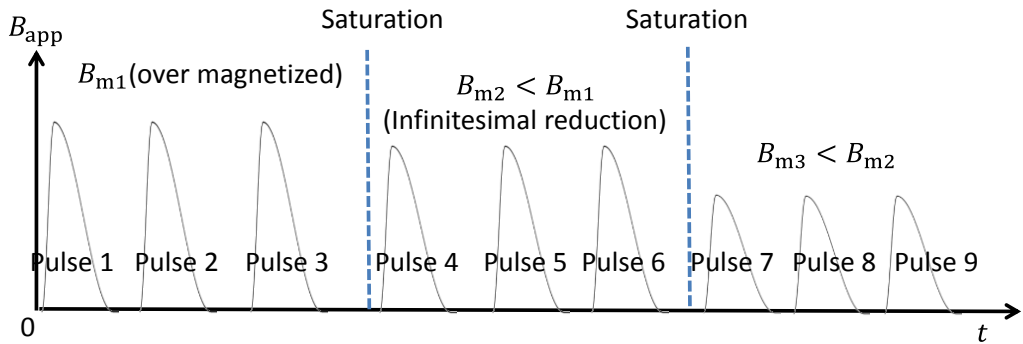

Fig. 14 Schematic diagram of the optimal applied pulse sequence deduced from the theoretical analysis.

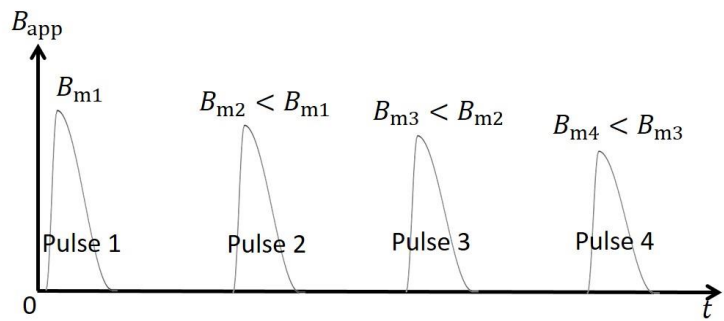

Fig. 15 Schematic diagram of the optimal applied pulse sequence with a finite pulse number.

## 3.5 Demonstration of optimal sequences

Based on the suggested optimal applied sequence in Section 3.4, the stack magnetized by typical pulse sequences are simulated and compared. The trapped field and flux with the amplitude of the each pulse is plotted in Fig. 16 and Fig. 18, respectively. "3.4/0.2" denotes a pulse sequence which starts from 3.4 T and reduces the amplitude by 0.2 T for each pulse. And so on, for other sequences. The trapped field and flux evolve along each line from the right to the left with each applied pulse.

Fig. 16 compares four pulse sequences starting with different starting amplitudes. The amplitudes reduce by 0.2 T for each successive pulse. Generally, the trapped field and flux improve with each applied pulse. The larger the starting amplitude, the larger the final trapped field and flux. When the starting amplitude is over 3 T, the final trapped field and flux are saturated. When the starting amplitude is below 3 T, the trapped field and flux are lower because it fails to make use of the central part of the stack. The final normalized critical current density $J/J_c$ of the four sequences are shown in Fig. 17. "3.4/0.2" and "3/0.2" have almost the same current density distribution; "2.6/0.2" and "2.2/0.2" have negative current density in the central part, which is under-magnetized.

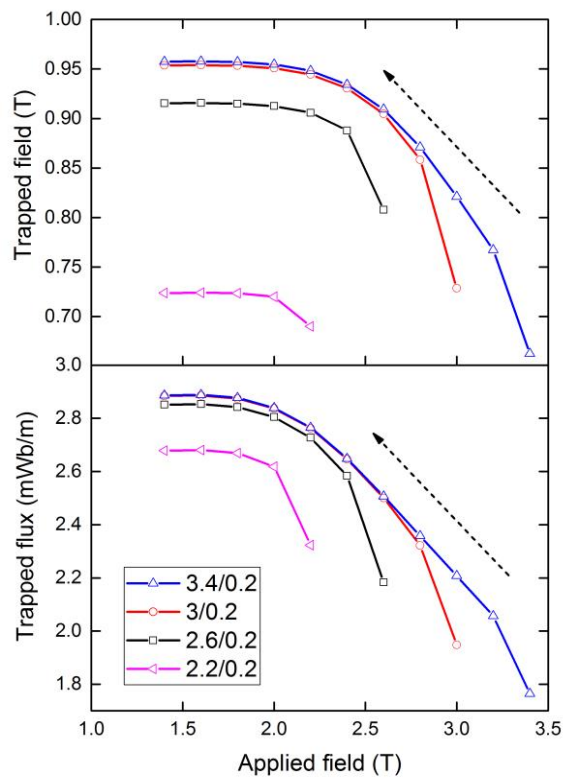

**Fig. 16 Demonstration of pulse sequences with different starting amplitudes. "3.4/0.2" denotes a pulse sequence which starts from 3.4 T and reduces the amplitude by 0.2 T for each pulse. And so on, for other sequences. The trapped field and flux evolve along each line from the right to the left with each applied pulse.**

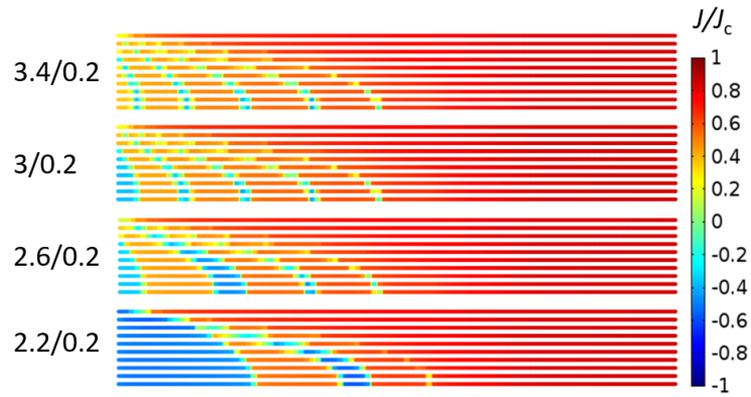

**Fig. 17 The final normalized critical current density $J/J_c$ of the four pulse sequences with different starting amplitudes corresponding to Fig. 16. Only a quarter of the cross-section of the stack is shown, as marked in Fig. 1. The thickness of the HTS layer is exaggerated for visualization.**

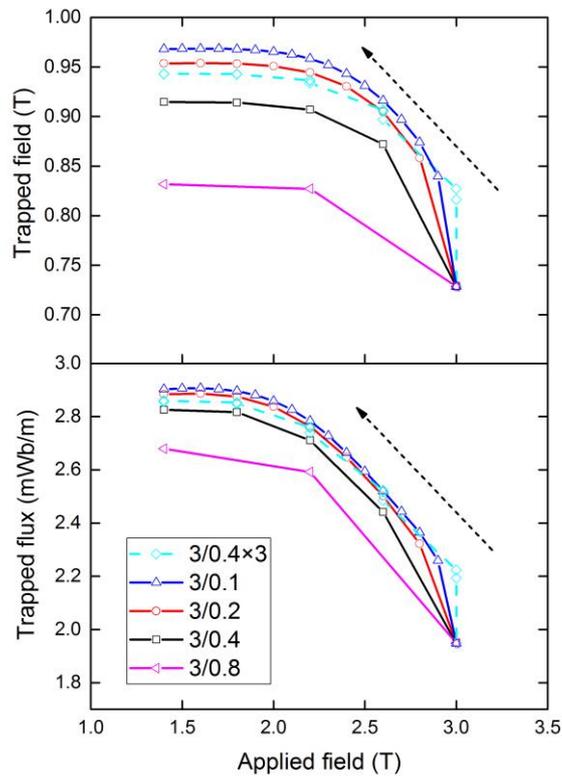

**Fig. 18 Demonstration of pulse sequences with different amplitude intervals. "3/0.1" denotes a pulse sequence which starts from 3 T and reduces the amplitude by 0.1 T for each pulse. And so on, for other sequences. "3/0.4×3" denotes that each amplitude repeats for three times in the sequence. The trapped field and flux evolve along each line from the right to the left with each applied pulse.**

Fig. 18 compares pulse sequences that start with the same amplitude (3 T) but have different amplitude intervals. In particular, "3/0.4×3" denotes that each amplitude repeats for three times and then reduce by 0.4 T in the sequence. "3/0.2" leads to more trapped field and flux than "3/0.4×3" with a smaller number of pulses, so it is not necessary to repeat the same amplitude considering that less pulses are preferred to save time and costs in practice as mentioned in Section 3.4.

Comparing the four pulses that start with 3 T and reduce the amplitude by different intervals, one can observe that the smaller the amplitude interval, the larger the final trapped field and flux. This is consistent with the deduction in Section 3.4. As shown in Fig. 19, with smaller amplitude intervals, the stack is magnetized by finer sections. This improvement will reach a saturation with the decrease of the amplitude interval.

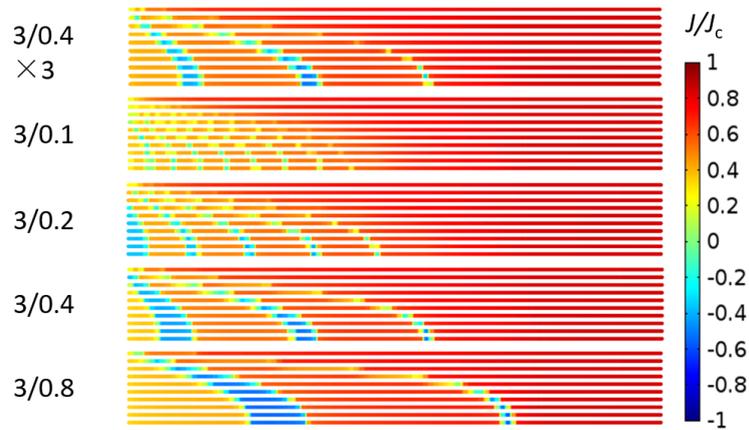

Fig. 19 The final normalized critical current density $J/J_c$ of the four pulse sequences with different amplitude intervals corresponding to Fig. 18. Only a quarter of the cross-section of the stack is shown, as marked in Fig. 1. The thickness of the HTS layer is exaggerated for visualization.

### 3.6 Experiments

In order to validate the simulation above, experiments were carried out on a stack of 20-layer CCs. The tape was 12 mm wide produced by SuperPower Inc. It was cut into 12*12 mm squares and tightly piled together. More information about the tapes can be found in [36]. The stack was magnetized by PFM at the working temperature of 30 K. The experimental setups are the same as those reported in [23]. The trapped field was measured 0.8 mm above the surface center of the stack 30 s after each applied pulse. After each pulse sequence, the stack was heated up to 95 K to ensure that the sample was "clean" of currents, so every tested sequence was started when there were no currents in the sample. Every pulse sequence was tested for three times with the same conditions on different dates to test for reproducibility and possible sample degradation. Corresponding to Fig. 16 and Fig. 18, the pulse sequences of different starting amplitudes or different amplitude intervals were tested, as shown in Fig. 20 and Fig. 21, respectively.

As shown in Fig. 20, the larger the starting amplitude, the larger the final trapped field, which is consistent with the simulation. The trapped fields are quite reproducible at large applied fields, but they start to deviate at about 3-4 T where the trapped fields are increasing rapidly. This may relate to the flux jumps which cannot be predicted by the numerical model in this work. The pulse sequences that start from 7 T and reduce the amplitude with different intervals are compared in Fig. 21. Consistent with the simulation, the smaller the amplitude interval, the larger the final trapped field. "7/0.2" shows a rather irregular pattern. It may relate to some instabilities that we cannot understand yet.

The deviations between the simulation and experiments include several aspects. The laminated structure and the material properties in the simulation are not the same for the sample in the experiments, which yields to the practical conditions. The simulation considers an infinitely long stack while the sample is a cuboid. The thermal condition in the experiments are difficult to judge. Despite these derivations, simulation and experiments lead to consistent optimal applied sequences for PFM as described in Section 3.4.

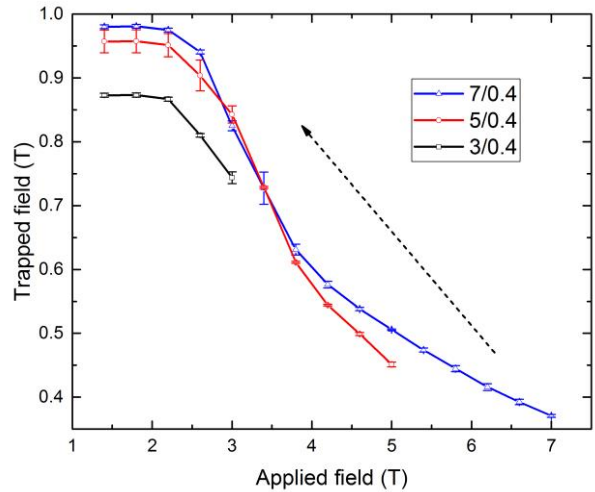

**Fig. 20** The experimental results of the trapped fields of a stack magnetized by pulse sequences with different starting amplitudes. "7/0.4" denotes a pulse sequence which starts from 7 T and reduces the amplitude by 0.4 T for each pulse.  And so on, for other sequences. The trapped field and flux evolve along each line from the right to the left with each applied pulse.

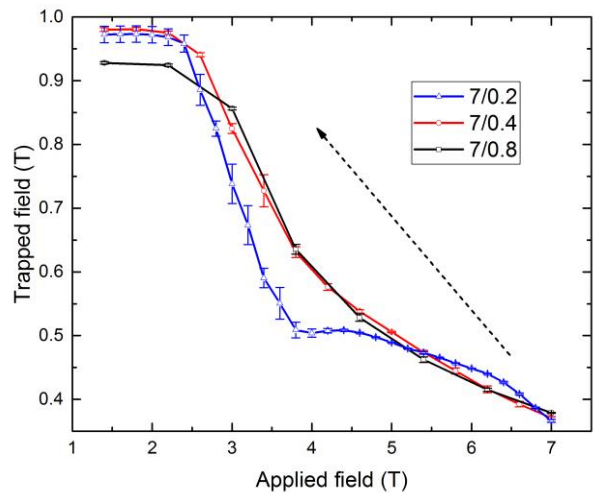

**Fig. 21** The experimental results of the trapped fields of a stack magnetized by pulse sequences with different amplitude intervals. "7/0.4" denotes a pulse sequence which starts from 7 T and reduces the amplitude by 0.4 T for each pulse.  And so on, for other sequences. The trapped field and flux evolve along each line from the right to the left with each applied pulse.

## 4. Conclusion

In this work, a systematic study on the stack of HTS (CCs) magnetized by PFM with multiple pulses at a fixed temperature was carried out numerically and experimentally. A 2D electromagnetic-thermal coupled model was used to simulate a stack of HTS coated conductors magnetized by PFM with successive pulses. A complete evolution tree of trapped field and flux following different pulse sequences was built. The evolution of the trapped field and flux with the amplitudes of successive pulses shows regular patterns as discussed in Section 3. Based on the derived patterns, the following optimal pulse sequence was suggested: first apply a pulse of large amplitude to over-magnetize the stack; then apply pulses of smaller amplitude (theoretically infinitesimal amplitude reduction one by one) to optimally magnetize the stack section by section. Demo pulse sequences are simulated and compared.

Correspondingly, a stack of CCs was magnetized by PFM with typical pulse sequences experimentally to validate the simulation. Larger starting amplitudes or smaller intervals between successive pulses produce more trapped field and flux, which is consistent with the simulation. Some sequences showed irregular patterns that may relate to flux jump. New models have to be developed to fully understand the process.

The derived patterns and strategy may apply to HTS bulks as well, given their similar non-linear *E-J* behaviors; however, the bulks, though simpler in geometry, have obvious grain boundaries and are much less uniform compared to tapes, which makes our conclusion not directly applicable. Nevertheless, similar pulse sequences as suggested in this work was proved to effectively improve the trapped field and flux of HTS bulks in experiments as reported in literature.

The work suggests the optimal sequences which significantly improve the trapped field and flux, but they still do not reach the maximum possible values (achieved by FC or ZFC). There is still room to make some further improvement by using complementary techniques. And how to reduce the time and costs of magnetization without sacrificing much of the performance is also a concern. Based on this work, a valuable further work will be to investigate the so-called multi-pulse technique with step-wise cooling (MPSC). It involves the optimization of both the applied sequences and the operating temperature. Combining all these techniques to improve the performance and lower the costs is crucial for *in situ* magnetization of HTS bulks or stacks to pave the way for practical and commercial use of HTS trapped field magnets.

**Reference**

[1] J. H. Durrell, *et al.*, "A trapped field of 17.6 T in melt-processed, bulk Gd-Ba-Cu-O reinforced with shrink-fit steel," *Supercond. Sci. Tech.*, vol. 27, pp. 082001, Aug. 2014.
[2] T. Tamegai, *et al.*, "Trapping a magnetic field of 7.9 T using a bulk magnet fabricated from stack of coated conductors," *Physica. C*, in press, available online: http://dx.doi.org/10.1016/j.physc.2016.02.006.
[3] A. Patel, *et al.*, "Trapped fields greater than 7 T in a 12mm square stack of commercial high-temperature superconducting tape," *Appl. Phys. Lett.*, vol. 102, pp. 102601, Mar. 2013.
[4] D. Zhou, *et al.*, "An overview of rotating machine systems with high-temperature bulk superconductors," *Supercond. Sci. Tech.*, vol. 25, pp. 103001, Aug. 2012.
[5] A. C. Day, *et al.*, "Design and testing of the HTS bearing for a 10 kWh flywheel system," *Supercond. Sci. Tech.*, vol. 15, pp. 838–841, Apr. 2002.
[6] T. Oka, *et al.*, "Application of HTS bulk magnet system to the magnetic separation techniques for water purification," *Physica. C*, vol. 468, pp. 2128–2132, Sep. 2008.
[7] S. B. Kim, *et al.*, "Study on Optimized Configuration of Stacked HTS Bulk Annuli for Compact NMR Application," *IEEE Trans. Appl. Supercond.*, vol. 21, no. 3, pp. 2080-2083, Jun. 2011.
[8] C. P. Bean, *et al.*, "Magnetization of high-field superconductors," *Rev. Mod. Phys.*, vol. 36, pp. 31–39, Jan. 1964.
[9] K. Rosseel, *et al.*, "Zylon-reinforced high magnetic field coils for the K.U. Leuven pulsed field laboratory," *Physica. B*, vol. 294-295, pp. 657–660, 2001.
[10] M. D. Ainslie, *et al.*, "Modelling of bulk superconductor magnetization," *Supercond. Sci. Tech.*, vol. 28, pp. 053002, Mar. 2015.
[11] H. Fujishiro, *et al.*, "Time evolution and spatial distribution of temperature in YBCO bulk superconductor after pulse field magnetizing," *Supercond. Sci. Tech.*, vol. 16, no.7, pp. 809, Jun. 2003.
[12] H. Fujishiro, *et al.*, "Generated heat during pulse field magnetizing for REBaCuO (RE = Gd, Sm, Y) bulk superconductors with different pinning abilities," *Supercond. Sci. Tech.*, vol. 18, no. 1, pp. 158, Nov. 2004.
[13] H. Fujishiro, *et al.*, "Estimation of temperature rise from trapped field gradient on superconducting bulk magnetized by multi-pulse technique," *Supercond. Sci. Tech.*, vol. 23, pp. 025013, Dec. 2009.
[14] H. Fujishiro, *et al.*, "Enhancement of trapped field and total trapped flux on GdBaCuO bulk by the MMPSC+IMRA method," *Supercond. Sci. Tech.*, vol. 22, pp. 095006, Aug. 2009.
[15] M. Sander, *et al.*, "Pulsed magnetization of HTS bulk parts at T<77 K," *Supercond. Sci. Tech.*, vol. 13, no. 6, pp. 841-845, Feb. 2000.
[16] H. Kamijo, *et al.*, "Repeated pulsed-field magnetization with temperature control in a high-T$_c$ bulk superconductor," *IEEE Trans. Appl. Supercond.*, vol. 11, no. 1, pp. 1816-1819, Mar. 2001.

[17] K. Murata, *et al.*, "Repulsed magnetization of a bulk superconductor by long and low field with temperature control," *Physica. C*, vol. 412-414, pp. 704–707, Oct. 2004.
[18] K. Kajikawa, *et al.*, "Finite element analysis of pulsed field magnetization process in a cylindrical bulk superconductor," *Physcia C*, vol. 468, pp. 1494-1497, 2008.
[19] H. Fujishiro, *et al.*, "Record-high trapped magnetic field by pulse field magnetization using GdBaCuO bulk superconductor," *Jpn. J. Appl. Phys.*, vol. 44, no. 37-41, pp. 1221-1224, Sep. 2005.
[20] H. Fujishiro, *et al.*, "Simulation of flux dynamics in a superconducting bulk magnetized by multi-pulse technique," *Physcia C*, vol. 471, pp. 889-892, 2011.
[21] S. Zou, et al., "Influence of parameters on the simulation of HTS bulks magnetized by pulsed field magnetization," *IEEE Trans. Appl. Supercond.*, vol. 26, no. 4, pp. 4702405, Jun. 2016
[22] K. Selva and G. Majkic, "Trapped magnetic field profiles of arrays of (Gd, Y) $Ba_2Cu_3O_x$ superconductor tape in different stacking configurations," *Supercond. Sci. Tech.*, vol. 26, pp. 115006, Aug. 2013.
[23] A. Patel, *et al.*, "Trapped fields up to 2 T in a 12 mm square stack of commercial superconducting tape using pulsed field magnetization," *Supercond. Sci. Tech.*, vol. 26, pp. 032001, Jan. 2013.
[24] A. Patel, *et al.*, "Pulsed-field magnetization of superconducting tape stacks for motor applications," *IEEE Trans. Appl. Supercond.*, vol. 25, no. 3, pp. 5203405, Jun. 2015.
[25] A. Patel, *et al.*, "Magnetic levitation using high temperature superconducting pancake coils as composite bulk cylinders," *Supercond. Sci. Tech.*, vol. 28, pp. 115007, Sep. 2015.
[26] S. Zou, et al., "Simulation of stacks of high-temperature superconducting coated conductors magnetized by pulsed field magnetization using controlled magnetic density distribution coils," *IEEE Trans. Appl. Supercond.*, vol. 26, no. 3, pp. 8200705, Apr. 2016.
[27] R. Brambilla, et al., "Development of an edge-element model for AC loss computation of high-temperature superconductors," Supercond. Sci. Tech., vol. 20, pp. 16-24, Nov. 2006.
[28] F. Grilli, et al., "Modeling high-temperature superconducting tapes by means of edge finite elements," IEEE Trans. Appl. Supercond., vol. 17, no. 2, pp. 3155-3158, Jun. 2007.
[29] V. Zermeño, et al., "A full 3D time-dependent electromagnetic model for Roebel cables," *Supercond. Sci. Tech.*, vol. 26, pp. 052001, May. 2013.
[30] M. Zhang, *et al.*, "Total AC loss study of 2G HTS coils for fully HTS machine applications," *Supercond. Sci. Tech.*, vol. 28, pp. 11501, Sep. 2015.
[31] J. Rhyner, "Magnetic properties and AC-losses of superconductors with power law current-voltage characteristics," *Phyica. C*, vol. 212, pp. 292-300, Jul. 1993.
[32] F. Grilli, et al.," Self-consistent modeling of the Ic of HTS Devices: How accurate do models really need to be?", *IEEE Trans. Appl. Supercond.*, vol. 24, no. 6, pp. 8000508, Dec. 2014.
[33] SuperPower, Inc., http://www.superpower-inc.com/system/files/2013_1105+LTHFSW+SuperPower+FINAL.pdf
[34] SuperPower, Inc., http://www.superpower-inc.com/content/2g-hts-wire
[35] V. Zermeno, et al., "Estimation of maximum possible trapped field in superconducting magnets in 2D and 3D," presented at the International Conference on Magnet Technology 24 (Presentation ID: 2PoBE_02), and the article is to be submitted to *Supercond. Sci. Tech.*.
[36] A. Baskys, et al., "Modeling of trapped fields by stacked (RE)BCO tape using angular transversal field dependence", IEEE Trans. Appl. Supercond., vol. 26, no. 3, pp. 6601004, Apr. 2016.